\magnification   \magstep1
\centerline {\bf  $ D_{00} $ Propagator in the Coulomb Gauge  }
\vskip 2cm
\centerline  { A. Andra\v si }
\centerline  {\it "Rudjer Bo\v skovi\' c" Institute, Zagreb, Croatia }
\vskip 4cm

\beginsection Abstract

The time-time component of the gluon propagator in the Coulomb gauge is believed to provide
a long-range confining force. We give the result, including finite parts, for the $ D_{00}$
propagator to order $ g^2 $ in the Coulomb gauge.
\vskip 1cm
PACS: 11.15.Bt; 11.10.Gh
\vskip 0.5cm
Keywords: Coulomb gauge; Renormalization
\vskip 5cm
Electronic address: aandrasi@rudjer.irb.hr

\vfill \eject

\beginsection 1. Introduction

The Coulomb gauge has many advantages over the covariant gauges. The propagators are closely related to the
polarization states of real spin-1 particles. There are no ghosts states. It is manifestly unitary.
Nevertheless, there are problems concerned with the convergence of the energy integrals [1]. The naive Coulomb gauge 
Feynman rules in non-Abelian gauge theory give rise to ambiguous integrals. At one-loop order and above there
are integrals like
$$ \int{{d^3P}\over{(2\pi)^2}}\int {{dp_0}\over{(2\pi)}}{{p_0}\over{p_0^2-P^2+i\eta}}\times{1\over{(P-K)^2}}.\eqno(1)$$
There is no regularization procedure for the energy divergence in $ p_0 $. These divergences have been studied in [2],[3]
where systematic cancellations have been found. However, ordinary ultra-violet divergences exist along with the above
energy divergences. No general proof exists that controls all divergences [4]. This means that a complete treatment of
the Coulomb gauge has not yet been given.
The Coulomb gauge has been extensively studied in the phase space formalism by D. Zwanziger [5] in the Euclidean space.
The ultra-violet divergent parts of the proper two-point functions have been calculated and found to observe the
Ward identities. In this paper we give the complete result for the time-time component of the gluon propagator to
order $ g^2 $ in Minkowski space and explore both limits $ k_0 \rightarrow \infty $ and $ k_0 \rightarrow 0 $ 
which are believed to give the colour-Coulomb potential.

\beginsection 2. The proper two-point functions

The zero-order propagators used in the Feynman rules in Minkowski space form a $ 7\times 7 $ matrix
 acting on $ A_1 $, $A_2 $, $A_3 $ ; $ A_0 $; $ E_1 $, 
$ E_2 $, $E_3 $ :
\vskip 0.5cm

\hrule height 1pt
\vskip 2mm
\settabs\+ $A_iMMMM$ &+ $-{T_{ij}}/{k^2}+\alpha{L_{ij}}/{K^2}MMMM $ & +$1/{K^2}+\alpha {k_0^2}/{(K^2)^2}MMMM$ &+$ -{ik_0T_{in}}/{k^2}MMMM$ \cr
\+  & $ A_j $ & $A_0$ & $ E_n $   \cr 
\vskip 2mm
\hrule height 1pt \vskip 2mm
\+ $A_i$ & $ -{T_{ij}}/{k^2}+\alpha{L_{ij}}/{K^2} $ & $ {\alpha k_0K_i}/{(K^2)^2} $ & $ -{ik_0T_{in}}/{k^2}$ \cr
\vskip 2mm
\+ $A_0$ & $ \alpha {k_0K_j}/{(K^2)^2}$ & $1/{K^2} +\alpha{k_0^2}/{(K^2)^2}$ & $i{K_n}/{K^2} $ \cr
\vskip 2mm
\+ $ E_m$ & $-i{k_0T_{mj}}/{k^2}$ & $i{K_m}/{K^2}$ & $-{T_{mn}K^2}/{k^2}$ \cr
\vskip 2mm
\hrule height 1pt
\vskip 5mm
where
$$ T_{ij}\equiv \delta_{ij}-L_{ij},{\hskip 1cm }  L_{ij}\equiv {K_iK_j}/{K^2}, $$
$$ k^2 = k_0^2-K^2. \eqno(2) $$
The Coulomb gauge propagators are obtained by setting $ \alpha = 0 $. The field $ E_i $ plays the role of the momentum
conjugate to $ A_i $. The constants used throughout the paper are
$$ \epsilon = 4-d  \eqno(3) $$
where $ d $ is the dimension of space-time and the coupling parameter is
$$ c={{ig^2}\over{16\pi^2}}C_G\delta_{ab}. \eqno(4) $$
There are two graphs which contribute to the $ A_0A_0 $ function. The graph shown in Fig.1 contributes$$ \Gamma_a^{A_0A_0} = c\{\Gamma({{\epsilon}\over 2})({{-k^2-i\eta}\over{\mu^2}})^{-{{\epsilon}\over 2}}\times
({1\over 2}k_0^2+{5\over 6}K^2+{{\epsilon}\over{12}}k^2+{{\epsilon}\over 6}k_0^2+{{17}\over{18}}\epsilon K^2) $$
$$ -2^{-\epsilon}({5\over 3}+{{28}\over 9}\epsilon)\Gamma({{\epsilon}\over 2})K^2({{K^2}\over{\mu^2}})^{-{{\epsilon}\over2}}
$$ $$ +{1\over 2}k^4\times D $$ 
$$+{{k^4}\over{2k_0K}}\ln{{k_0+K-i\eta}\over{k_0-K+i\eta}}\times \ln{{K^2}\over{(-k^2-i\eta)}} $$
$$-k_0^2\ln{{K^2}\over{(-k^2-i\eta)}}-2(\ln 2 -1)k_0^2 \} \eqno(5) $$
The graph in Fig.2 gives
$$ \Gamma_b^{A_0A_0}=c\{\Gamma({{\epsilon}\over 2})({{-k^2-i\eta}\over{\mu^2}})^{-{{\epsilon}\over 2}}\times({1\over 3}K^2
-{1\over 2}k^2-{{\epsilon}\over 4}k^2+{{11}\over{18}}\epsilon K^2) $$ $$ -{1\over 3}\Gamma({{\epsilon}\over 2})K^2
({{K^2}\over{{\mu}^2}})^{-{{\epsilon}\over 2}} -K^2({{10}\over 9}-{{2\ln 2}\over 3}) $$
$$ +k^2k_0^2\times D $$ 
$$+{{k_0k^2}\over{K}}\ln{{k_0+K-i\eta}\over{k_0-K+i\eta}}\times \ln{{K^2}\over{(-k^2-i\eta)}} $$
$$ -(2k^2+K^2)\ln{{K^2}\over{(-k^2-i\eta)}} -2(\ln 2-1)(2k^2+K^2)\} \eqno(6) $$
where the non-rational structure $ D $ which appears in the results in the integral form is
$$ D=\int_{0}^{1}dx {{x^{-{1\over 2}}}\over{k_0^2-x(K^2-i\eta)}}\ln(1-x). \eqno(7) $$

\beginsection  In the region $ k_0>K $

$$ D={1\over{k_0K}}\{Li_2({{k_0-K+i\eta}\over{k_0+K-i\eta}}) -Li_2({{k_0+K-i\eta}\over{k_0-K+i\eta}})  $$
$$ +\ln{{k_0+K-i\eta}\over{k_0-K+i\eta}}\times \ln{{k^2+i\eta}\over{K^2}}-i\pi \ln{{k_0+K-i\eta}\over{k_0-K+i\eta}}\}.
\eqno(8) $$

\beginsection  In the region $ K>k_0 $

$$ D={1\over{k_0K}}\{Li_2({{K-k_0-i\eta}\over{K+k_0-i\eta}})-Li_2({{K+k_0-i\eta}\over{K-k_0-i\eta}})  $$
$$+\ln{{K+k_0-i\eta}\over{K-k_0-i\eta}}\times \ln({{-k^2-i\eta}\over{k_0^2}})+i\pi \ln{{K+k_0-i\eta}\over{K-k_0-i\eta}}\}
$$ $$ -{2\over{k_0K}}[Li_2(-{{k_0}\over{K-i\eta}})-Li_2({{k_0}\over{K-i\eta}})] +{{i\pi}\over{k_0K}}
\ln{{K^2}\over{(-k^2-i\eta)}}  \eqno(9) $$
where
$$ Li_2(x)=-\int_{0}^{x}{{\ln (1-z)}\over {z}} dz \eqno(10)  $$
is the Spence function. The two expressions for $ D $ in (8) and (9) are connected as analytic continuations
of each other.

\beginsection 3. The time-time propagator

 The propagators in the Coulomb gauge are obtained by inverting the $ 7\times 7 $ matrix of the
proper two point functions. The complete matrix will be given in a future publication. However, the time-time
component of the gluon propagator to order $ g^2 $ involves only two more proper functions.
$$ D^{A_0A_0}={1\over{K^4}}[\Gamma^{A_0A_0}+iK_n\Gamma^{A_0E_n}] $$
$$+{{iK_m}\over{K^4}}[\Gamma^{E_mA_0}+iK_n\Gamma^{E_mE_n}]  \eqno(11) $$
To one loop order these proper functions are
$$ \Gamma^{A_0E_i}=c({{K^2}\over{\mu^2}})^{-{{\epsilon}\over 2}}\{{4\over 3}-2^{-\epsilon}
\Gamma({{\epsilon}\over 2})({2\over 3}+{{13\epsilon}\over 9})\}\times(2iK_i) \eqno(12) $$  and
$$ \Gamma^{E_iE_j}=-2c({{K^2}\over{\mu^2}})^{-{{\epsilon}\over 2}}\{2^{-\epsilon}({2\over 3}+
{{13\epsilon}\over 9})\Gamma({{\epsilon}\over 2})\delta_{ij} -{4\over 3}{{K_iK_j}\over{K^2}}\} \eqno(13) $$
leading to
$$ D^{A_0A_0}= c(K^2)^{-2}\{{{11}\over 3}\Gamma({{\epsilon}\over 2})K^2 -{5\over 3}K^2\ln{{(-k^2-i\eta)}\over{\mu^2}}
-2K^2\ln{{K^2}\over{\mu^2}} $$ $$ +{1\over 2}k^2(k^2+2k_0^2)\times D $$
$$+{{k^2}\over{2k_0K}}(k^2+2k_0^2)\ln{{k_0+K-i\eta}\over{k_0-K+i\eta}}\times \ln{{K^2}\over{(-k^2-i\eta)}} $$
$$-(3k_0^2-K^2)\ln{{K^2}\over{(-k^2-i\eta)}}-(6k_0^2+2K^2)\ln 2 +6k_0^2+{{31}\over 9}K^2\} \eqno(14) $$
The UV divergent part of $ D^{A_0A_0} $ agrees with [5], [6] and [7] showing the antiscreening nature of the QCD vacuum.

\beginsection 4. Discussion

There are two interesting limits of eq.(14). The Zwanziger limit [5] $ k_0 \rightarrow \infty $ is
$$ \lim_{k_0 \rightarrow \infty} D^{A_0A_0}(k_0,K) = {c\over{K^2}}\{{{11}\over 3}\Gamma({{\epsilon}\over 2})
-{{11}\over 3}\ln{{K^2}\over{\mu^2}}-i\pi-{{28}\over3}\ln 2 +{{103}\over 9} -2\ln{{K}\over{k_0}}\} \eqno(15) $$
and it is not independent of $ k_0 $. The limit $ k_0 \rightarrow 0 $ is naturally related to the definition of the
quark-antiquark potential.
It follows from considering a rectangular Wilson loop with sides of length $ T $ in the time direction (where
$ T \rightarrow \infty $ ) and $ L $ in the space direction. In the Coulomb gauge the main contribution comes from
the $ D_{00} $ component of the propagator (where $ k_0 \rightarrow 0 $) attached to the two time-like sides.
In this limit ( $ T\rightarrow \infty $) we expect the contribution to tend to zero for any graph with one end
attached to either of the spacelike sides. This leaves graphs connected to the timelike sides at each end. Such graphs
involve the $ D_{00} $ propagator, which we have calculated, crossing from one side to the other.
The contribution of such a loop gives the potential between two very heavy quarks separated by the distance $ L $,
$$ V(L) = -4\pi g_B^2 \int d^{3-\epsilon}K e^{iK\cdot L} D_{00}(0,K) \eqno(16) $$
where $ g_B $ is the bare coupling constant.
The limit $ k_0 \rightarrow 0 $ of (14) including the zero'th order propagator is
$$ \lim_{k_0 \rightarrow 0} D^{A_0A_0}(k_0,K) ={i\over{K^2}}\{1+c'[{{11}\over 3}\Gamma({{\epsilon}\over 2})
-{{11}\over 3}\ln{{K^2}\over{\mu^2}} +{{31}\over 9}]\} \eqno(17) $$ 
where $$ c' ={{g^2}\over{16\pi^2}}C_G \delta_{ab}. \eqno(18) $$
The bare coupling constant $ g_B $ is related to the renormalized coupling constant $ g_R $
$$ g_B = (1-{{11}\over 3}{{c'}\over{\epsilon}})g_R {\mu}^{{{\epsilon}\over 2}}. \eqno(19) $$
Inserting (17) and (19) into (16)  leads to the quark-antiquark potential
$$ V(L) =-2\pi^2g_R^2(\mu){1\over L}\{1+{{31}\over 9}c'+{{11}\over 3}c'\gamma +{{11}\over 3} c'\ln(\mu L)^2\} \eqno(20)$$
where $ \gamma $ is the Euler's constant and $ g_R(\mu) $ the running coupling constant. If we assume the relation
$$ L\times \mu =1 \eqno(21) $$
$g_R(\mu) $ becomes $ L $ dependent. We suppose that the exact $ g_R({1\over L}) $ tends to zero as $ L\rightarrow 0 $.
Also  $ g_R({1\over L}) \rightarrow \infty $ for $ L \rightarrow \infty $. 

\beginsection  Acknowledgment

A.A. wishes to thank Prof. J. C. Taylor for the invaluable help and advice which made this work possible.
The author is grateful to the Arany J\' anos Foundation of Hungarian Academy of Sciences for financial support.
This work was supported by the Ministry of Science and Technology of the Republic of Croatia under Contract
No. 0098003.

\beginsection  References

[1] H. Cheng and E. C. Tsai, Phys. Lett. {\bf B176}(1986)130;

\noindent
[2] P. Doust and J. C. Taylor, Phys. Lett. {\bf 197}(1987)232;

\noindent
[3] P. Doust, Ann. of Phys. {\bf 177}(1987)169;

\noindent
[4] J. C. Taylor, in Physical and Nonstandard Gauges, Proceedings, Vienna, Austria 1989,
\line{\hskip 0.5cm P.Gaigg, W. Kummer and M. Schweda (Eds.); \hfil} 

\noindent
[5] D. Zwanziger, Nucl. Phys. {\bf B518}(1998)237;

\noindent
[6] J. Frenkel and J. C. Taylor, Nucl. Phys. {\bf B109}(1976)439;

\noindent
[7] T. D. Lee, in Particle Physics and Introduction to Field Theory, Harwood Academic \break
\line{\hskip 0.5cm Publishers, 1981, secs. 18.5 and 18.6; \hfil}

\vskip 1cm

\beginsection Figure Captions

Fig.1. The time-time component of the gluon self-energy to order $ g^2 $. The dotted lines represent the 
instantaneous Coulomb field $A_0$. The continuous line is the $ E_i $ field conjugate to the transverse field $ A_i$.
The propagators inside the loop are the $ E_iA_j $ transitions specific to the Coulomb gauge.
\vskip 0.5cm
\noindent
Fig.2. Self-energy graph to order $ g^2 $. The dotted line is the $ A_0 $ field, the dashed line is the
transverse propagator and the solid line is the $ E_iE_j $ propagator. 
\vskip 0.5cm
\noindent
Fig.3 The dashed line is the transverse gluon propagator which attaches to the spacelike sides of length $ L $
of the Wilson loop. The timelike sides are of length $ T $. In the limit $ T \rightarrow \infty $ the contribution
of this loop vanishes.
\vskip 0.5cm
\noindent
Fig.4 The dotted line is the zero'th order Coulomb propagator attached to the timelike sides of length $ T $.
In the limit $ T \rightarrow \infty $ this loop gives the potential between two very heavy quarks which just
sit at the positions $ (0,0,0) $ and $ (L,0,0) $ for all the time. To zero'th order propagator this just gives
the Coulomb potential $ {1\over L } $.

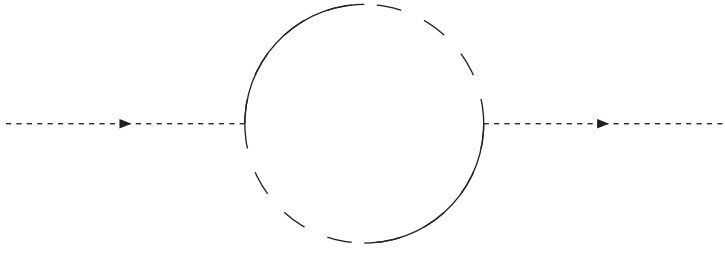
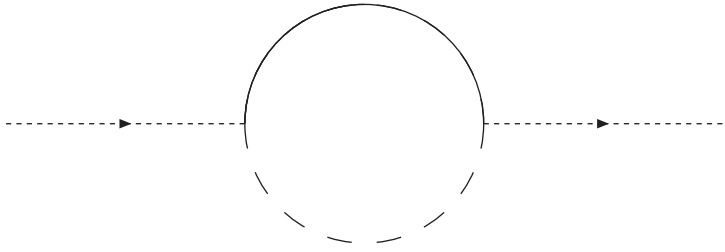
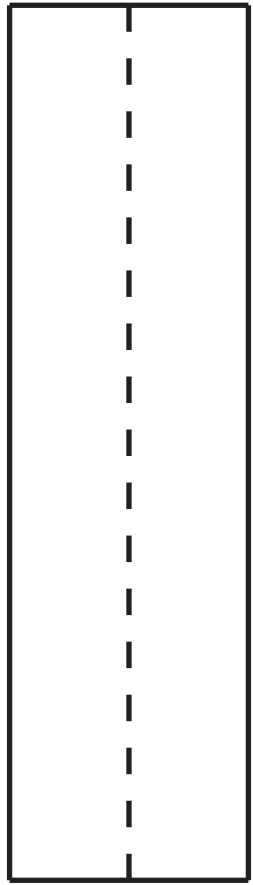
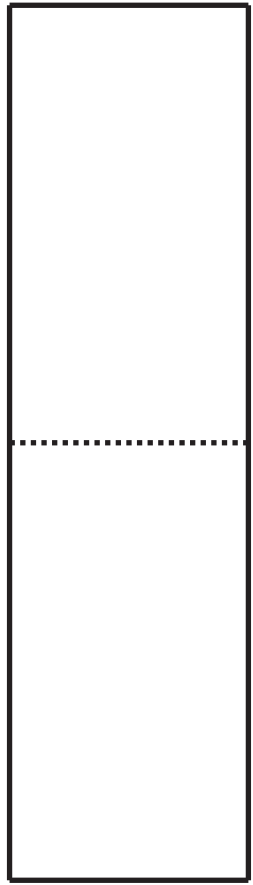
\bye